# Cluster based Key Management in Wireless Sensor Networks

Anvesh Aileni

Computer Science Department

Oklahoma State University, Stillwater, OK – 74078

anvesh@okstate.edu

**ABSTRACT**

Wireless sensor networks consist of sensor nodes with limited computational and communication capabilities. In this paper, the whole network of sensor nodes is divided into clusters based on their physical locations. In addition, efficient ways of key distribution among the nodes within the cluster and among controllers of each cluster are discussed. Also, inter and intra cluster communications are presented in detail. The security of the entire network through efficient key management by taking into consideration the network's power capabilities is discussed. A graphical representation of the simulation on the scheme is also presented.

## 1. INTRODUCTION

A sensor network consists of spatially distributed nodes with limited computational and communication capabilities. These are used in varied applications such as monitoring environmental conditions, military and other industrial purposes. The nodes which are distributed spatially have to communicate with each other in a secured manner since the data associated with these nodes may be confidential. Efficient cryptography techniques are to be used for communication among the nodes. Use of public key cryptography is omitted here keeping in mind the computational limitations of sensor nodes [1]. Also, using a single key for the entire network may be a compromise on the security of the system as a threat on even one node would bring the system down.

Due to the storage limitations of the sensors nodes, all unique keys cannot be stored. Consider a total of *m* nodes out of which two nodes share a unique key. This will result in a total of $\left(\frac{m(m-1)}{2}\right)$ keys in the network with each node storing $(m-1)$ keys. This is very high owing to the storage constraints of a node if *m* is large. Also, considering the ad-hoc nature of the sensor networks, some nodes will continue to be added to the network over time. Thus efficient key management has to be performed in order to store keys in nodes such that the network security is not compromised and also the total number of keys stored in the node is reduced.

Symmetric key pre distribution is an approach used earlier but our proposed scheme deals with dividing the entire network into clusters based on their topography. Eschenauer and Gligor [2] proposed a "key ring" scheme using random graphs wherein, from a large pool of keys *S,* few keys were selected and stored in the node such that two neighbors share at least a key with certain probability *p*. Each node carries a subset of keys called the key ring of size *m* from *S* and finds common set of keys among subset of any two randomly chosen key rings. This scheme requires three main steps namely, key pre-distribution, key discovery and path key identification. In the key discovery phase every node sends a message to all other nodes. The neighbors upon receiving the message determine which key are shared with which node. Also a path is established among nodes if they do not share a key and this is done in path key identification. This populates the entire network and this process may have to be repeated every time a node is added or a node is replaced.

The key ring approach requires that data be routed over many nodes before it reaches its destination. This indicates that there is an underlying connection between routing and security and that implicit security technique [3],[4],[5] may be used to route data via multiple channels, thus spreading the vulnerability over several nodes and communication channels making an adversary's task harder. There is also the question of node compromise which may be countered by using tamper-resistance hardware in each node [6]. Further, overlay architecture has been used to achieve balance between number of keys per node and routing complexity [13].

In this paper, a modification of the key ring approach will be presented that divides the set of nodes into several sub networks. Each of these sub networks that are defined using a clustering algorithm, has its own sub controller.

## 2. PROPOSED SCHEME

The network is divided into clusters, not necessarily of same size. Each divided cluster will have a sub controller in addition to a main controller which belongs to all the clusters. The next phase after clustering is to load the keys into the memory of the nodes and these keys stored in the memory is called a key set. Keys are also to be distributed among the sub controllers for their communication, as all the sub controllers may not be in the communication range. After key distribution is done, the deployment phase begins and the communication among the nodes is established. Figure 1 gives an idea of how the entire network is divided into 4 clusters.

In a cluster, say of *n* nodes, there are $\left(\frac{n(n-1)}{2}\right)$ pairs of keys. Each sub controller has to find a way of populating the key set of each sensor node. Also, main controller will do the work similar to that of sub controller in distributing the keys among sub controllers. Once the network is divided into clusters, it is assumed that each node interacts only with its peers within the cluster and there is no direct interaction with nodes outside that cluster.

Other assumptions are that each sub controller is within the range of its own cluster and main controller can reach all sub controllers. Also, sub controllers have more communication and computation capabilities than other nodes in the network and likewise the main controller has more computational capabilities than sub controllers and nodes. One more important assumption is that the main controller can reach any sub controller in the network but the opposite may not be true.

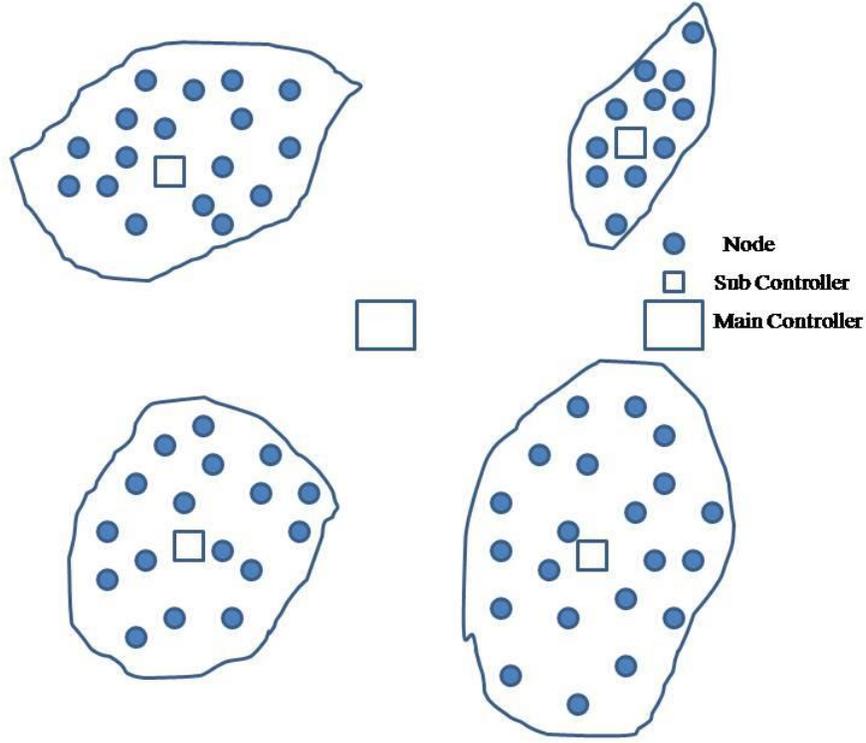

Figure 1. A network divided into various clusters

## 2.1 Clustering

Grouping of nodes in a network into clusters will be done by the *k*-means++ clustering algorithm [7]. In *k*-means++ clustering algorithm, the number of clusters is fixed *a priori*. We assume that the number of clusters to be chosen is *k* and this choice is based on the network size and geographical positions of each node. Consequently, we will have *k* sub controllers monitoring nodes in their respective clusters.

According to *k*-means++ clustering algorithm, initially *k* means are to be identified, one for each cluster. For this, as explained in *k*-means++ approach, following steps are to be followed,

1. Choose an initial centre *x*, from the set of all node points say µ, uniformly at random.
2. Choose next mean from the set and name it *y* with probability $\frac{D(y)^2}{\sum_{x \in \mu} D(x)^2}$ proportional to $D(x)^2$
3. Repeat step 2 until all the *k* centers are identified.

Once these centers are identified, these would be the means of the clusters and each node distance from all the means is calculated and is associated to the nearest mean, this process is called binding. This process is repeated up to a stage where all the positions of nodes in the network are visited at least once. After this, $k$ new means will be calculated. Once the positions of $k$ new means are identified, binding is done again. This process is repeated until there is no change in the location of means. The main goal is to minimize the sum of squares of distance within each cluster given by (1)

$$\sum_{i=1}^{k} \sum_{x_j \in S_i} ||x_j - m_j||^2 \longrightarrow (1)$$

Where $m_1, m_2, m_3 ..... m_k$ are the means of the clusters initially defined.

$$S_i^t = \{ x_j : || x_j - m_i^t || \leq || x_j - m_{i*}^t || \text{ for all } i* = 1, 2, \ldots k\}$$

$x_j$ is the relative distance of $(x,y)$ in the network from origin.

$m_i$ is the relative distance of mean (x,y) from origin.

Updating the means:

$$m_i^{t+1} = \frac{1}{|S_i^t|} \sum_{x_j \in S_i^t} x_j$$

$t$ – Ranges from 1, 2... till the means does not change

$j$ – 1, 2 ..... n.

$i$ – 1, 2 ..... n.

Thus, the $k$ means locations will be identified and these will remain as locations for controllers of those clusters.

### 2.2 Key Distribution

As the communication can be both inter and intra cluster, the key distribution should also be considered for two cases, one among nodes and other among sub controllers.

#### 2.2.1 Key Distribution among nodes

In this scheme, the birthday problem is used in distributing keys among nodes in the network [8],[9],[10]. According to the birthday problem we have to find number of people with certain probability, so that at least two among them share a common birthday.

Let the probability with which we have to find number of people $n$ in a room to be $p(n)$ and number of days to be $d$. So according to the birthday problem

$$p(n) = 1 - \frac{n!\binom{d}{n}}{d^n}$$

The equation can be approximated to

$$p(n) = 1 - \left(1 - \frac{1}{d}\right)^{C(n,2)}$$

We can draw an analogy to relate to the present scenario as follows,

1. Total number of nodes in a cluster can be compared to number of days.
2. Number of people in a room can be associated with number of nodes sharing keys with each node in a cluster.

Assume total number of nodes to be *m* in the network and the network is divided into *k* clusters. Choose randomly a cluster with *n* nodes with probability *p(s)*, given by

$$p(s) = 1 - \left(1 - \frac{1}{n}\right)^{C(s,2)}$$

Here, *s* is the number of nodes with which any node shares a key.

As the value of *p* changes, *s* value changes and that give us the total number of nodes where two nodes share a unique key. Consider a random node *i* in the cluster and randomly choose *s* nodes which share a unique key with node *i* in the cluster. Out of $\left(\frac{n(n-1)}{2}\right)$ total keys, those nodes with which node *i* shares a key are stored in the memory. This process is repeated till all the nodes in the cluster store keys.

### 2.2.2. Key Distribution among Sub Controllers

Not every sub controller can interact with all other sub controllers due to its limited communication capabilities. So, a key distribution should also be present among the sub controller to have an effective communication. To accomplish this, all those sub controllers which are within the communication range are to be identified first. Then, the same method as that of the nodes can be employed here.

Let *l* be the number of clusters identified within a communication range of distance of *d*. From the birthday problem, we can identify the number of sub controllers sharing keys among themselves. Say a sub controller *si* has to share keys with *ss* sub controllers within its range out of the total *l* number of sub controllers. The probability *p(ss)* can be determined as,

$$p(ss) = 1 - \left(1 - \frac{1}{l}\right)^{C(ss,2)}$$

Here we get the value of *ss* i.e., total number of sub controllers with which sub controller *si,* shares a key with. Those *ss* sub controllers are chosen randomly from *l* which are in the range of that sub controller under consideration.

**2.3 Path Identification**

After key distribution phase, path has to be established among different nodes which may either belong to same cluster or to a different cluster. There are two ways in which this can be done.

1. Path among sub controllers
2. Path among nodes.

**2.3.1 Path among Sub Controllers**

When there is a inter cluster communication, there has to be a path that exists between sub controllers. For this purpose Dijkstra's shortest path algorithm [11],[12] is used. All the weights are assumed to be equal to 1.

According to the shortest path algorithm the following steps are to be followed in order to identify the path from one sub controller to other.

1. Initially assign a sub controller and identify all those sub controllers which can interact with it.
2. Assign all the sub controllers status as *not visited*.
3. When all the nodes neighboring the current nodes are visited, mark the current node visited.
4. If by the $3^{rd}$ step all the nodes are visited search ends there and if not, mark the next node as current node and repeat the process from step 2.

Thus, by this procedure all the paths between different sub controllers are identified and stored in their respective memory.

**2.3.2 Path among Nodes**

Each path has to be identified from a source node to a destination node. Instead of populating the network at the beginning, the path is established between two nodes when it is required. That is, when a message has to be transferred from one node to another for the first time, a path will be identified using the key sets and this path is remembered for a next time use. This way, path is identified between different nodes and stored in sub controller. The two cases that are to be dealt are:

**1.** Both source node and destination node belongs to same cluster
**2.** Both of them belong to different clusters.

### 2.3.2.1. Case 1

Assume a path has to be identified between nodes *i* and *j* which are in the same cluster. Node *i* sends message, if *i* and *j* share a key, the message is broadcasted directly, otherwise, i checks for node *n*, where

$\{n \mid \text{if } n \in \{\text{key set of } i\} \text{ and } \forall n \text{ where } d_{kj} < d_{xj}, x = \{\text{key set of } i\}\}$

$d_{nj}$ is the distance between node *n* and *j*

When the message is broadcasted to *n*, the next nearest node of *j* is searched and this process is repeated till there is a path directly from *n* to *j*.

Consider an example of a path that has to be identified from node say *c* to node *r*, by the procedure discussed we can identify *n*'s to be nodes *o*, *m*, *p*, *u*, *t*, *e*, and thus the path from *c* to *p* is *c-o-m-p-u-t-e-r*. This path identified will be stored and if for the next time there is a communication required from *c* to *p*, the same path is followed instead of identifying the path again in the similar fashion.

### 2.3.2.2. Case 2

When a node has to interact with node in another cluster, sub controller communication is also needed along with node communication. Thus, when a node in one cluster has to communicate with a node in another cluster, that node sends a message to its sub controller. This sub controller based on other sub controller's information sends a message to the destination cluster's sub controller through the path which has already been defined. Thus, after reaching the sub controller of destination cluster, it will send a message to the first node of its cluster and then the process as discussed in case 1 is repeated. The path thus established here again will be stored.

## 3. SIMULATIONS

Simulations are done to investigate the results using different network size. Simulations assume a network of 4000 nodes in the network and 4, 8 and 12 clusters in each case.

Number of hops passed on an average in 4 clusters is shown in the graph of Figure 2. This is the simulation done for a network of 4000 nodes and 4 clusters. In a cluster of 1000 nodes with 499500 keys we take only key set of size which varies from 15 to 204, 15 being for probability 0.1 and 204 for 0.999999999. But total number of keys in the network is 7998000. At 0.1 probabilities number of hops on an average it takes to reach destination is 28.7 while it is 2 at 0.999999999.

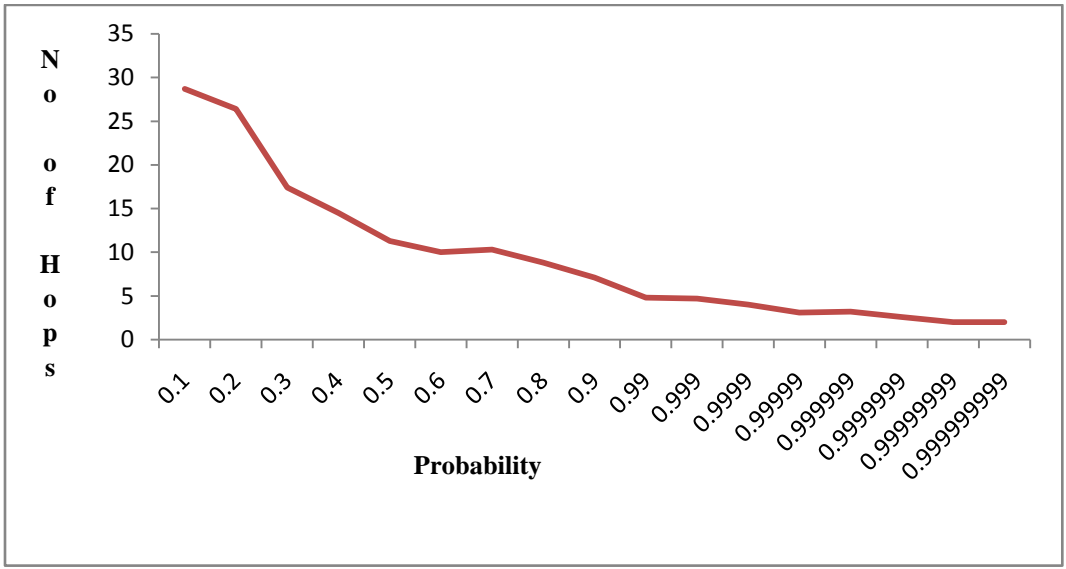

Figure 2. No of hops on an avg vs probability for 4 clusters

Figure 3 presents the case of 8 clusters for a network of size 4000 nodes. In a cluster of 500 nodes with 124750 keys we take only key set of size which varies from 10 to 144, 10 being for probability 0.1 and 144 for 0.999999999. The total number of keys in the whole network is 7998000. At 0.1 probabilities the number of hops taken to reach the destination is 19.3 while it is 1.9 at 0.999999999.

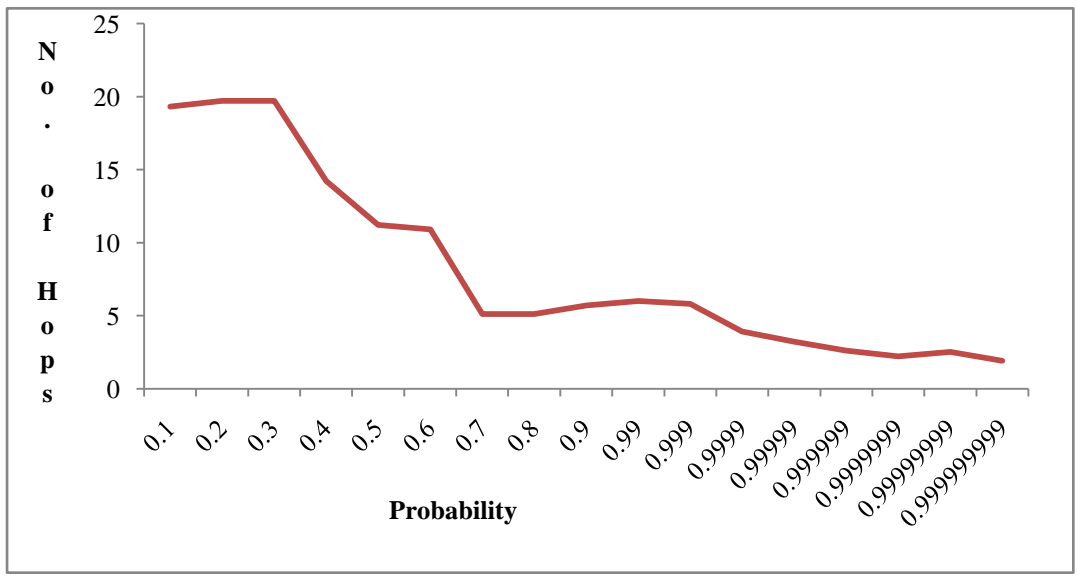

Figure 3. No of hops on an avg vs probability for 8 clusters

### 3.1. Comparison of hops and key set size for clusters

The graph of Figure 4 shows how the key set size is reduced when number of clusters is increased at different probabilities even though the network size is same. At probability of say 0.5, the key set size for node is 37, 35 and 21 when the network is divided into clusters of 4, 8 and 12 respectively.

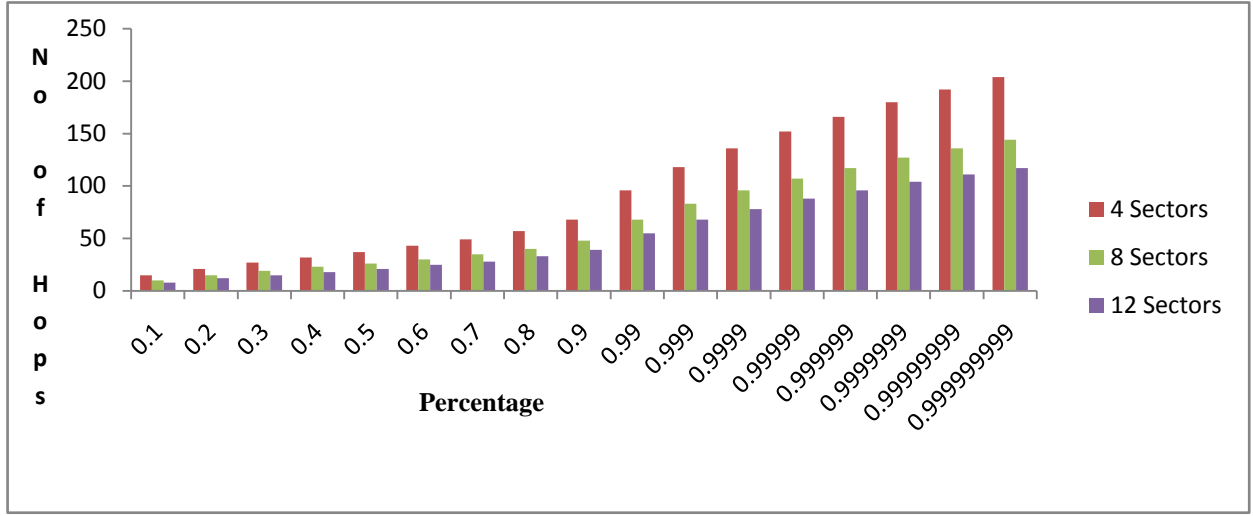

Figure 4: Number of Hops versus Percentage

Similar observations may be made for the number of hops taken to travel from the source node to the destination node. The graph of Figure 5 shows us how the number of hops gets reduced when number of clusters is increased at different probabilities even though the network size is the same. At probability say 0.5, the number of hops it takes is 11.3, 11.2 and 7.8 when the network is divided into clusters of 4, 8 and 12 respectively.

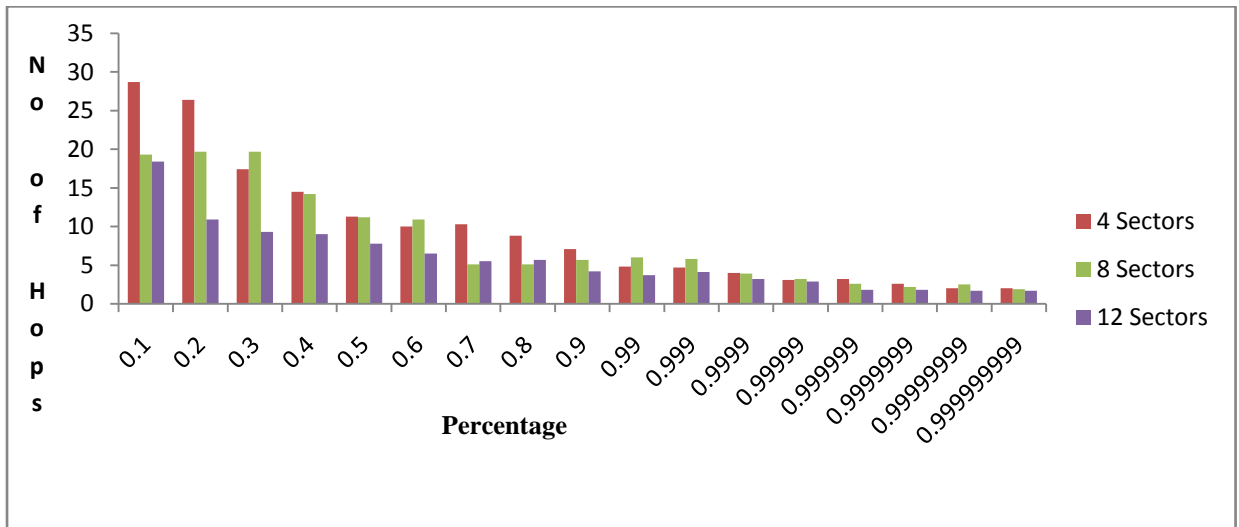

Figure 5: Number of Hops versus Percentage

## 4. CONCLUSION

The scheme presented in this paper is an effective protocol for dividing the network into clusters and for distributing keys among them. This method is efficient when the nodes in the network are divided randomly and can be clustered easily rather than the nodes when distributed in a uniform fashion. Simulations were run for the proposed design and the results are presented in graphs. These results show that the performance in terms of number of hops and number of keys stored in a node improves as the number of clusters increases.